\documentclass[useAMS,usegraphicx,usenatbib]{mn2e}

\def\ltsima{$\; \buildrel < \over \sim \;$}
\def\simlt{\lower.5ex\hbox{\ltsima}}
\def\gtsima{$\; \buildrel > \over \sim \;$}
\def\simgt{\lower.5ex\hbox{\gtsima}}
\def\kpc{{\rm\,kpc}}

\newcommand{\apj}{ Astrophys. J.}

\newcommand{\apjs}{Astrophys. J. Suppl.}
\newcommand{\aj}{Astron. J.}
\newcommand{\mnras}{Mon. Not. R. Astron. Soc.}
\newcommand{\asa}{A\&A}

\newcommand{\pasp}{Pub. Astron. Soc. Pac.}

\makeatletter
\newcommand\ion[2]{#1$\;${\small\rmfamily\@Roman{#2}}\relax}%
\makeatother

\def\s{\ifmmode \widetilde \else \~\fi}
\def\={\overline}

\def\spose#1{\hbox to 0pt{#1\hss}}

\def\lta{\mathrel{\spose{\lower 3pt\hbox{$\mathchar"218$}}
     \raise 2.0pt\hbox{$\mathchar"13C$}}}
\def\gta{\mathrel{\spose{\lower 3pt\hbox{$\mathchar"218$}}
     \raise 2.0pt\hbox{$\mathchar"13E$}}}
\def\Dt{\spose{\raise 1.5ex\hbox{\hskip3pt$\mathchar"201$}}}    
\def\dt{\spose{\raise 1.0ex\hbox{\hskip2pt$\mathchar"201$}}}    

\def\dotsfill{\leaders\hbox to 1em{\hss.\hss}\hfill}

\title{A faint extended cluster in the outskirts of NGC~5128: evidence of a low mass accretion}

\author[Mouhcine et al.]{M.~Mouhcine$^1$, W.~E.~Harris$^2$, R.~Ibata$^3$, M. Rejkuba$^4$ \\
$^1$Astrophysics Research Institute, Liverpool John Moores University, 
    Twelve Quays House, Egerton Wharf, Birkenhead, CH41 1LD, UK \\
$^2$Department of Physics \& Astronomy, McMaster University,
      Hamilton ON L8S 4M1, Canada \\
$^3$Observatoire Astronomique de Strasbourg (UMR 7550),
    11, rue de l'Universit\'e, 67000 Strasbourg, France \\
$^4$European Southern Observatory, Karl-Schwarzschild-Strasse 2, 
D-85748 Garching, Germany  }

\date{Accepted ?. Received ?; in original form ?}

\pagerange{\pageref{firstpage}--\pageref{lastpage}}
\pubyear{2007}

\begin{document}

\maketitle

\label{firstpage}

\begin{abstract}

We report the discovery of an extended globular cluster in a halo field in Centaurus A 
(NGC~5128), situated $\sim 38\kpc$ from the centre of that galaxy, imaged with the 
Advanced Camera for Surveys on board the Hubble Space Telescope. At the distance of the 
galaxy, the half-light radius of the cluster is ${\rm r_h \sim 17\,pc}$, placing it among the largest 
globular clusters known. The faint absolute magnitude of the star cluster, $M_{V,\circ}=-5.2$, 
and its large size render this object somewhat different from the population of extended 
globular clusters previously reported, making it the first firm detection in the outskirts of a giant 
galaxy of an analogue of the faint, diffuse globular clusters present in the outer halo of the Milky 
Way. The colour-magnitude diagram of the cluster, covering approximately the brightest four 
magnitudes of the red giant branch, is consistent with an ancient, i.e., ${\rm \ga 8\,Gyr}$, 
intermediate-metallicity, i.e., ${\rm [M/H]\sim -1.0 dex}$, stellar population. We also report the 
detection of a second, even fainter cluster candidate which would have ${\rm r_h \sim 9\,pc}$, 
and $M_{V,\circ}=-3.4$ if it is at the distance of NGC~5128. The properties of the extended 
globular cluster and the diffuse stellar populations in its close vicinity suggest that they are 
part of a low mass accretion in the outer regions of NGC~5128.

\end{abstract}


\section{Introduction}
\label{intro}

\footnotetext[1]{This work was based on observations with the NASA/ESA Hubble Space Telescope, 
obtained at the Space Telescope Science Institute, which is operated by the Association of 
Universities for Research in Astronomy, Inc.,under NASA contract NAS 5-26555. }

The generic properties of globular clusters, which are generally considered
to be free of significant amounts of dark matter, are
different from those of the dark-matter-dominated dwarf galaxies. These classes of objects
were found for a long time to occupy different regions of parameter space in 
terms of, e.g., scale size, luminosity, and internal velocity 
dispersion \citep[e.g.][and references therein]{mateo98}.

New systematic surveys for globular clusters and dwarf satellites over the last few years 
have generated a shift in 
our understanding of the connection between those two classes of objects. On one hand, several
of the newly discovered dwarf satellites around the Milky Way \citep[e.g.][]{willman05,belokurov07} 
and Andromeda \citep[e.g.][]{zucker04,martin06,irwin08} exhibit luminosities and/or sizes that 
are strikingly comparable to those of globular clusters. On the other hand,  
luminous star clusters with large sizes have been discovered around, e.g., M31 \citep{huxor05} and
M33 \citep{stonkute08}. In contrast, in the Milky Way 
globular cluster system, most of the unusually extended star clusters are the Palomar-type objects
that lie preferentially at large 
Galactocentric distances and are predominantly faint \citep[e.g.][]{vandenbergh04}. The extended 
luminous star clusters are found to resemble the classical globular clusters in terms of their 
stellar population properties, i.e., they are generally old and metal-poor, albeit with combined 
structures and luminosities unlike those observed for any other globular clusters in the Local 
Group or beyond \citep{mackey06}. They are found to occupy the gap in parameter space between 
classical (compact) globular clusters and dwarf spheroidal galaxies \citep{huxor05}.

Globular clusters are typically found in much larger numbers within giant elliptical galaxies 
than in spirals such as the Local Group large galaxies. At a distance of 3.8 Mpc 
\citep{rejkuba04,harris09}, NGC~5128 is the closest giant elliptical. Its globular cluster system, the 
largest of any galaxy within ${\rm \sim 15\,Mpc}$, has been studied photometrically and 
spectroscopically for many years
\citep[for only the most recent such work and references to earlier papers, see, e.g.][]{harris04,woodley07,gomez07,mcl08,woodley09}, 
and at present a total of 605 clusters have been individually identified. Extended (and bright) 
globular clusters are found to represent a tiny fraction of its overall star cluster population. 
\citet{gomez07} indicate that globular clusters with half-light radii exeeding 8\,pc make up 
a few percent of the their spectroscopically confirmed globular clusters. These extended 
globular clusters have luminosities comparable to those of extended clusters in the outskirts 
of M31 \citep[see Fig.~8 of][]{gomez06}), brighter than most Galactic globular clusters of similar 
sizes. The most massive clusters in NGC~5128 are found to follow a mass-size relation, and to have
higher mass-to-light ratios on average than clusters with masses below ${\rm 10^{6} M_{\odot}}$
\citep{rejkuba07}.

In this contribution, we report the discovery of two faint, extended globular cluster candidates
in the outskirts of NGC~5128, one of which is a high-probability old globular cluster based on 
direct colour-magnitude photometry.  
The dataset used here was initially obtained to study the resolved stellar populations in the 
remote outskirts of NGC~5128, allowing us to investigate the properties of the neighborhood of 
the extended faint star cluster to shed light on its origin. The layout of this paper is as follows: 
Section \S~\ref{data} briefly describes the data set, while Section \S~\ref{results} presents the 
properties of the reported halo globular cluster. In \S~\ref{disc} we will discuss the present work 
and its implications. \S~\ref{summary} gives the summary.

Throughout the paper we use an intrinsic distance modulus for NGC 5128 of$(m-M)_{\circ}=27.90$ 
following the recent synthesis of five standard-candle distance calibrations given by \citep{harris09}. 
The foreground reddening toward the field was estimated to be $E(B-V)=0.11$ from the all-sky map 
of \citet{schlegel98}. We have neglected the effect of any possible differential reddening across 
the targeted field and along the line of sight from within NGC 5128 itself, since the target field 
is one located in its outer halo. The differential reddening due to the Galactic foreground across 
the $3.4'$ width of an HST/ACS field at this Galactic latitude ($b = 19^o$) is also likely to be 
negligible.

\section{Data}
\label{data}

The imaging data we use here is the single very deep pointing obtained by \citet{rejkuba05} 
with the Advanced Camera for Surveys (ACS/WFC) camera on the Hubble Space Telescope (HST)
in program GO-9373. This field, imaged in $F606W$ (wide $V$) and $F814W$ ($I$), is located 
at a linear projected distance of approximately $\sim 38\kpc$ South of the NGC 5128 center. 
The target field was chosen to avoid any known surface brightness anomalies such as jet-induced 
star-forming regions \citep{mould00,rejkuba02}, shells and arcs \citep{malin83}, and dust lanes 
\citep{stickel04}. The total exposure consisted of of 12 full orbits in each filter, reaching 
$F606W \sim 30$, $F814W \sim 29$ and making this material by far the deepest photometric probe 
into the stellar population of NGC 5128. In terms of limiting absolute magnitude, it is deep 
enough to resolve horizontal-branch stars in the halo and to show the structure of the 
``red clump'' (the core-helium-burning stars in the colour-magnitude diagram). The observations 
and reduction techniques are fully described in \citet{rejkuba05} and we therefore only briefly 
summarize this information here.

The photometry of \citet{rejkuba05} was measured with the DAOPHOT suite of codes \citet{stetson1994}. 
From this we select stars with high quality measurements, including sharpness parameter $< 2$, 
goodness of fit $\chi^{2} < 3$, and $\sigma_{\rm F606W} < 0.3$. We further limited the catalogue 
to stars with $-0.5 < (V-I)_{\circ} < 4.0$.

For an old simple stellar population, redder stars are more metal-rich than bluer ones, and the 
width of the red giant branch (RGB) at a given luminosity is a very direct indicator of the spread 
in the stellar metallicity. The analysis of the properties of the age-sensitive features of the 
asymptotic giant branch bump and the red clump of the stellar populations in the outskirts of 
NGC~5128 indicates that they are predominantly old. While we cannot rule out a modest age spread 
in these halo stars \citep[see][]{rejkuba05}, the colour spread of the RGB stars observed in the 
data is due primarily to a spread in metallicity at an old mean age and can be used to derive the 
halo metallicity distribution function (MDF) \citep[e.g.][and references therein]{harris99,harris00,harris02,mouhcine05}. 
To estimate the metallicities of the survey stars, we repeat the procedure explained 
in \citet{mouhcine07}, interpolating between the models of \citet{vandenberg06} of $\alpha$-enhanced 
red giant branch stars of mass $0.8\,M_{\odot}$. The models span ${\rm -2.314 < [Fe/H] < -0.397}$ 
approximately in steps of 0.1 dex, and are complemented at the metal rich end by two models with 
${\rm [Fe/H] = 0.0}$ and ${\rm [Fe/H] = +0.4}$ (see \citealt{mouhcine05} and \citealt{harris00} 
for more details on the interpolation procedure and the calibration of the track grid). 
Stars outside of the colour-magnitude range covered by these RGB tracks were flagged and not used 
in subsequent analysis (i.e., no extrapolation beyond the validity of the models was attempted). 
The metallicity of stars on the red clump cannot be interpolated from RGB tracks; we therefore 
imposed a faint cut-off at $M_I < -0.75$ to retain only the brighter RGB stars, which also have the 
advantage of being clearly brighter than the completeness limits of the data. This limit corresponds 
to $I_{\circ} < 27.17$, or $I < 27.38$, where the $I$-band measurement uncertainty is $\sigma_I=0.13$. 
An additional quality cut of $\sigma_V=0.13$ was imposed to ensure that the $V$-band measurements 
were also of good quality.

\section{Results}
\label{results}

\subsection{A new faint globular cluster}

\begin{figure}
\includegraphics[clip=,width=0.45\textwidth]{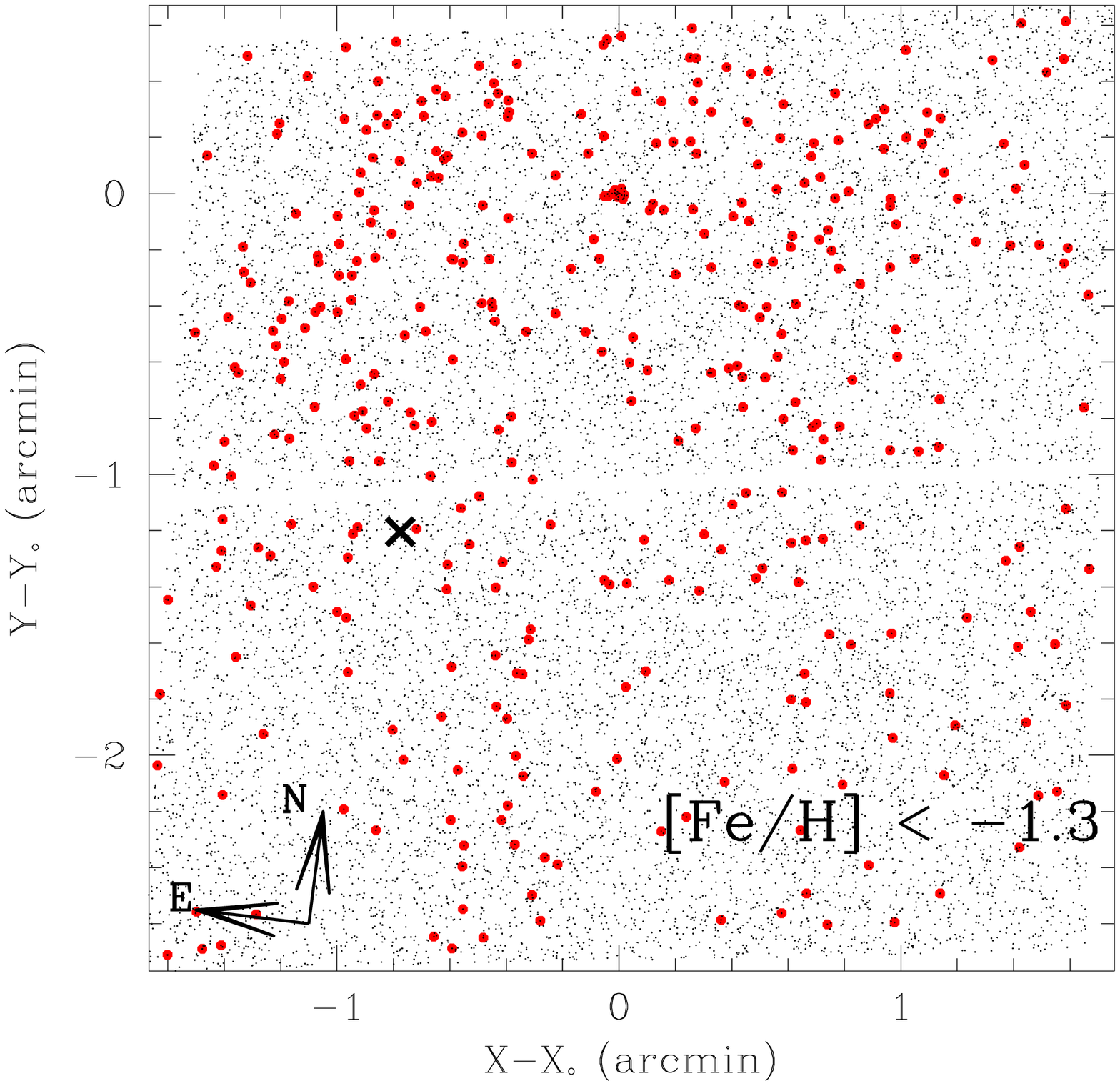}
\caption{Spatial distribution of stellar sources classified as RGB candidates over the observed
field. Large solid points show the distribution of metal-poor, i.e., [Fe/H] $< -1.3$, stars. 
The small compact clump of metal-poor stars at top center shows the position of the candidate 
star cluster G0606. The cross indicates the position of the second star cluster candidate G0607. 
The position scales are normalized to $(0,0)$ at the cluster candidate G0606.}
\label{xy_map}
\end{figure}

\begin{figure*}
\includegraphics[angle=0]{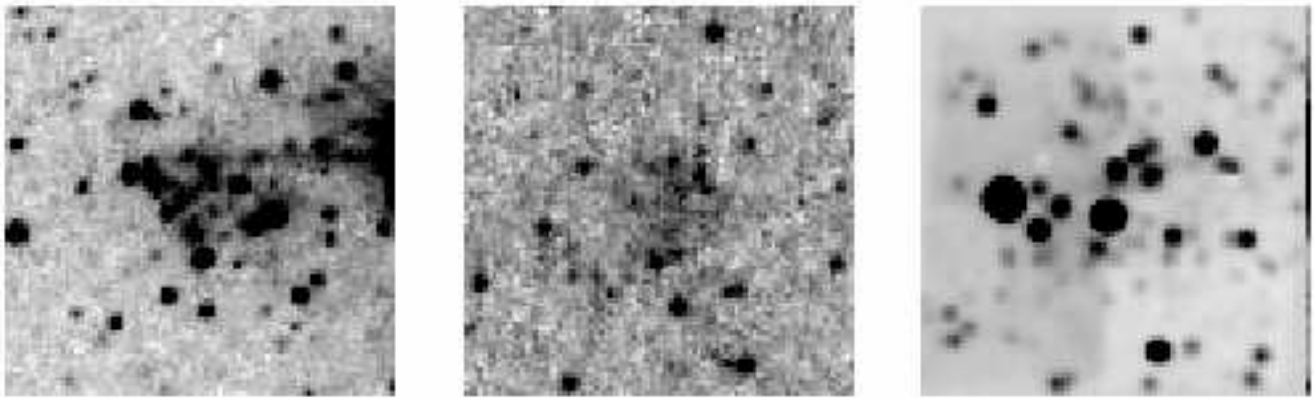}
\caption{Thumbnail images in $V$ of the two cluster candidate, GC0606 (left panel) and GC0607 
(center panel) (see the text for the adopted numbering scheme).  Field size of each thumbnail
is 4 arcsec across, equivalent 
to 75 pc at the distance of NGC 5128.   Note the very bright star just off the right side
of the frame for GC0606.  The \emph{right panel} shows a thumbnail image for 
comparison of the Milky Way outer-halo cluster Palomar 14, drawn from the HST/ACS archives.  
Its luminosity and effective radius are similar to our two cluster candidates.
To compensate for the $50\times$ nearer distance of Pal 14, its image was resampled with a 
Gaussian convolution and then block-averaged to the same linear size.}
\label{images_gc}
\end{figure*}

The overwhelming majority of stars in this outer-halo pointing are the rather uniformly
distributed halo field stars.  However, the long dynamical timescales of regions this
far out (38 kpc and above) allow the possibility that identifiable substructures, such as
extended globular clusters (GCs) or faint satellite dwarfs might be present as well.
An efficient way for revealing the presence of these stellar (sub-)structures in
the presence of heavy field contamination, especially when integrating stellar populations 
along a line of sight, is to investigate differences in the spatial distribution of stars 
with different metallicities \citep{ibata07,ibata09}. Figure\,\ref{xy_map} shows the spatial 
distribution of RGB stars over the field. The sample of all the field stars is shown as 
small dots, while \emph{metal-poor} RGB stars, i.e., those with photometrically determined 
abundances [Fe/H] $< -1.3$ dex, are shown as larger solid dots. The most striking feature 
in the stellar spatial distribution is a small, compact  clump of metal-poor RGB stars near 
the west side of the frame (see Fig.~\ref{xy_map}). No such clumps are found for the more 
chemically evolved stellar populations. 

To shed light on the nature of the identified concentration of metal-poor stars, we show on the left 
panel of Figure\,\ref{images_gc} a zoomed thumbnail image from the ACS $F606W$ stacked frame. 
The object clearly has the morphology of a faint, extended star cluster, more diffuse in structure 
than the classical globular clusters that have previously been observed in NGC~5128 (see, 
e.g. the sample images in \citet{harris06}). Following the homogeneous catalogue numbering
system for NGC 5128 GCs defined by \citet{woodley07} and continued in \citet{woodley09}, 
we name this new GC candidate GC0606. 

This new object strongly resembles the Palomar-type clusters found in the Milky Way outer halo.
For more direct visual comparison, the right panel of Figure\,\ref{images_gc} shows a thumbnail 
image of the Milky Way outer-halo cluster Palomar 14 (which is $\sim 70$\,kpc from the Sun), 
drawn from the HST/ACS archives. To compensate for the difference in distances between NGC~5128 
and Pal 14, i.e., a factor of 50, the Pal 14 image was resampled with a Gaussian convolution 
and then block-averaged to the same linear size. Pal 14 has a scale size (half-light radius 24 pc) 
and luminosity ($M_V = -4.7$) that are similar to our NGC 5128 candidates (see below),
demonstrating that these faint, extended GCs can be discovered readily with this technique.

\begin{figure}
\includegraphics[clip=,width=0.45\textwidth]{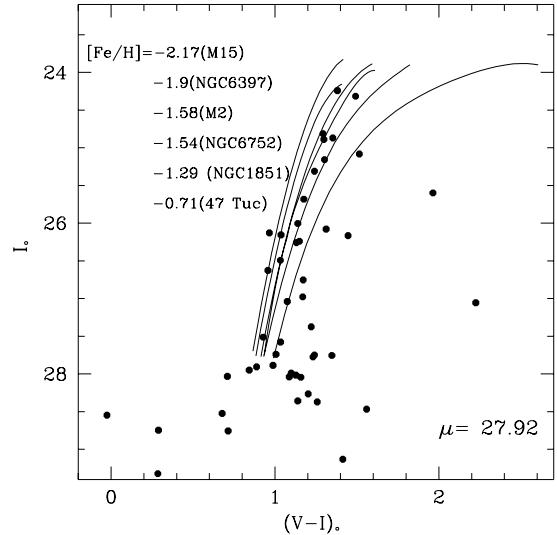}
\caption{The colour-magnitude diagram of stars lying within a radius of 2 arcsec from the centre
of the cluster candidate GC0606. Shown also as solid lines are the red giant sequences of the 
indicated Galactic globular clusters from \citet{dacosta90} assuming an intrinsic distance 
modulus for NGC~5128 of 27.9. A sequence of red giant stars is clearly visible. }
\label{gc_cmd_fidu}
\end{figure}

\begin{figure}
\includegraphics[clip=,width=0.45\textwidth]{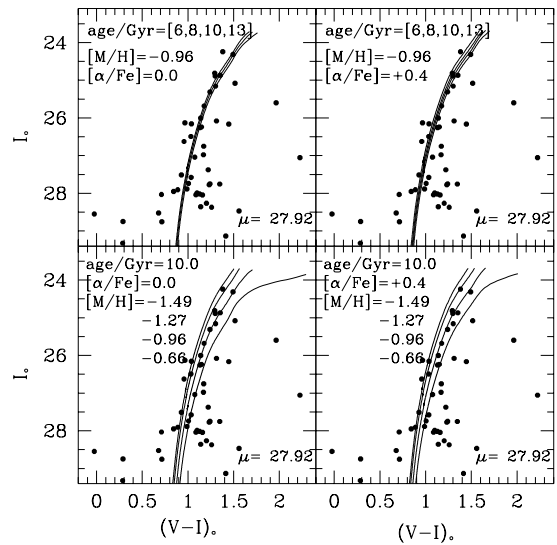}
\caption{The cluster colour-magnitude diagram of the cluster candidate GC0606 compared 
to red giant branches from scaled-solar (the left panels) and $\alpha$-enhanced BaSTI 
isochrones with ${\rm [\alpha/Fe]=+0.4}$ (the right panels). The upper panels show the 
effect of varying the age at a fixed overall metallicity, while the lower panels show the effect 
of varying the overall metallicity at a fixed age. }
\label{gc_cmd_basti}
\end{figure}

A powerful tool to investigate the nature of the candidate is the colour-magnitude diagram 
(CMD) of its stars. A classical globular cluster with a half-light radius of a few parsecs would 
subtend only a small angle:  a typical half-light radius of 3\,pc converts 
to 3 ACS pixels, not much larger than the 2-px Full Width at Half Maximum of the stellar point 
spread function. The outer parts of normal, luminous GCs will have enough stars that a CMD can 
be obtained; this was first done for NGC 5128 by \citet{harris98} for the cluster C44=GC0227, 
an inner-halo cluster which turned out to have moderately low metallicity from its RGB locus.
The deep WFPC2/PC1 $(V,I)$ images used for their study reached to $I(lim) \simeq 27$, not as 
deep as our ACS data.

In the case of faint and diffuse star clusters, the situation is less desperate. 
Even though the total RGB population within the cluster GC0606 is low, 
the crowding is not extreme and the ACS images contain a number of resolved members for which 
the PSF fitting procedure allows accurate photometric measurements. In order to place constraints 
on the stellar population properties of GC0606, we show in Fig.\,\ref{gc_cmd_fidu} the CMD of 
all stellar objects detected within a circular region with a radius of two arcseconds (40 px) 
around the star cluster. Fiducials of RGB sequences for Galactic globular clusters spanning a 
wide range of metallicities from \citet{dacosta90} and shifted to the distance of NGC~5128 are 
overplotted. The brightest stars in the cluster are nicely aligned with the red-giant sequences 
at intermediate metallicity, between the fiducial sequences for ${\rm [Fe/H]\sim -1.3}$ and 
$-1.55$, particularly over the upper two magnitudes where the photometry is of the highest 
quality ($\sigma_{V,I} \la 0.05$). This CMD not only confirms
the identity of GC0606 as an old globular cluster, but also places it comfortably within the 
metal-poor sub-population of the bimodal metallicity distribution function of the globular 
cluster system of NGC~5128 that peaks at ${\rm [Fe/H]\sim-1.5}$ \citep[e.g][]{harris04,beasley08}.

To strengthen the constraints on the properties of the star cluster, we compare the stellar 
photometry to the BaSTI theoretical isochrones for both scaled-solar and $\alpha$-enhanced 
compositions \citep{pietrinferni04,pietrinferni06}. Fig.\,\ref{gc_cmd_basti} shows aligned 
Solar-scaled (left panels) and $\alpha$-enhanced (right panels) isochrones on the star 
cluster CMD. The upper panels show comparison with isochrones with ages ranging from 6\,Gyr 
to 13\,Gyr for an overall metallicity of ${\rm [M/H]=-0.96}$, while the lower panels show 
10\,Gyr isochrones with metallicities ranging from -0.66 to -1.5. 
Considering first the Solar-scaled isochrones, those with a metallicity 
${\rm [M/H]=-0.96}$ and ages older than 6\,Gyr are found to provide the best fit along 
the majority of the RGB. 
The BaSTI theoretical isochrones also give a metallicity estimate in good agreement with 
our empirical value of ${\rm [Fe/H]\sim -1.3}$, keeping in mind that metal-poor GCs typically 
exhibit enhancement in $\alpha$-elements of approximately ${\rm [\alpha/Fe] \sim +0.3}$ 
\citep[see e.g.][]{carney96}.

The comparison of the star cluster CMD to the BaSTI isochrones with an $\alpha$-enhanced 
composition, i.e., ${\rm [\alpha/Fe] = +0.4}$, indicates that the isochrones with 
${\rm [M/H]=-0.96}$ and ages older than ${\rm \sim\,6\,Gyr}$ provide again the best fit to the 
cluster RGB. Once again, this is in excellent agreement with the metallicity estimate obtained 
by comparing the star cluster to Galactic cluster fiducials.


\subsection{Structural Parameters}

To characterize further the globular cluster, we measured its overall structural parameters. 
A surface brightness profile of the globular cluster was generated by carrying out aperture 
photometry of the cluster on both the $V$ and $I$ master images. We derived the profile from 
\emph{daophot/apphot} direct concentric-aperture photometry, using the flux differences 
between apertures of adjacent radii to derive the surface brightness in successive annuli. 
A problem we faced in this analysis was the presence of a nearby very bright star, which 
prevented the profile from being measured further out than 16 px radius ($0.8''$) in $I$ 
and 26 px ($13''$) in $V$. 

\begin{figure}
\includegraphics[angle=0,width=0.5\textwidth]{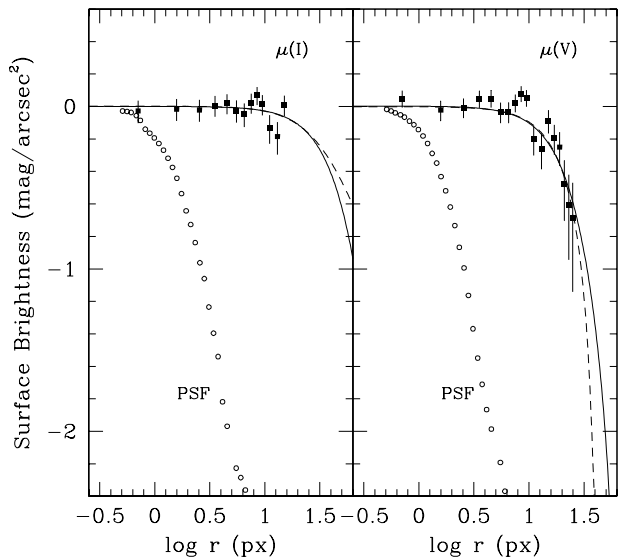}
\caption{Surface brightness profile for the brighter candidate, GC0606.  The measurements were 
made through direct concentric-aperture photometry as described in the text. Solid dots show the 
measurements in concentric annuli with internal error bars, while open circles show the fiducial 
profile for the stellar PSF.  The best-fit King (1966) model is shown as the solid line, while 
the King (1962) model fit is shown as the dashed line.  Left and right panels show the $I-$band
and $V-$band results separately; the vertical axis give the surface brightness normalized to the 
central value $\mu_0$. }
\label{brightprofile}
\end{figure}

The profile fitting code of \citet{mcl08} was then used to fit \citet{king62} and \citet{king66} 
profiles, numerically convolved with the stellar PSF as defined from nearby bright, uncrowded
stars on the frames. The best-fit results for both model profiles are shown 
in Figure \ref{brightprofile}. As is evident from the figure, the cluster is very much more 
extended in characteristic size than the PSF, so the fitting results are insensitive to the 
details of the PSF or any small changes it might have across the field.  
The center coordinates are no more certain than $\pm 2$ px ($0.1''$) on each axis.

The cluster is almost entirely dominated by a large, flat central core.  In the sense defined 
by the King models, the cluster has an extremely low formal value of the central concentration 
$c = {\rm log} (r_t/r_c)$ and central dimensionless potential $W_0$, making it similar to 
the lowest-concentration globular clusters known. The half-light radius $r_h$ was 
determined by numerically integrating the PSF-deconvolved King profile solutions out to large 
radius and then finding the projected radius enclosing half the total luminosity. 
Lastly, we estimated central surface brightness, 
which is also insensitive to the details of either the profile fit or the PSF because the 
cluster has essentially constant surface brightness within 5 px. All the integrated magnitudes 
and central surface brightness values are internally uncertain to at least $\pm 0.1$ magnitude 
and perhaps more. 

The compiled measured parameters are shown in Table \ref{list_gc_candidates}. 
The apparent magnitudes and colours have been converted to luminosities and intrinsic colours 
assuming $(m-M)_I = 28.10$ and $E_{V-I} = 0.14$, and conversion of half-light radii to parsecs 
assumes $d = 3.8$ Mpc as stated earlier. It is evident that the cluster is just as diffuse 
and low-luminosity as its first visual impression gives. 

\begin{table}
\caption{Physical Parameters of Faint Extended Globular Cluster Candidates}
\label{list_gc_candidates}
\begin{tabular}{lcc}
\hline
Parameter & GC0606 & GC0607 \\
\hline
$\alpha$ (J2000) & 13:25:12.40 & 13:25:19.78 \\
$\delta$ (J2000) & -43:35:15.44 & -43:34:40.30 \\
$I$(tot) & 21.9 & 24.0 \\
$(V-I)$ & 1.15 & 0.84 \\
$\mu_0(I)$ (mag arcsec$^{-2}$) & 23.0 & 24.1  \\
$r_h$ (arcsec) & $0.80 \pm 0.03$ & $0.47 \pm 0.03$ \\
$M_I$ & -6.2 & -4.1 \\
$M_V$ & -5.2 & -3.4 \\
$r_h$ (pc) & 14.7 & 8.7 \\
$L_V/L_{\odot}$ & $10.3 \times 10^3$ & $1.9 \times 10^3$ \\
\end{tabular}
\end{table}

\begin{figure*}
\includegraphics[angle=0,width=0.45\textwidth]{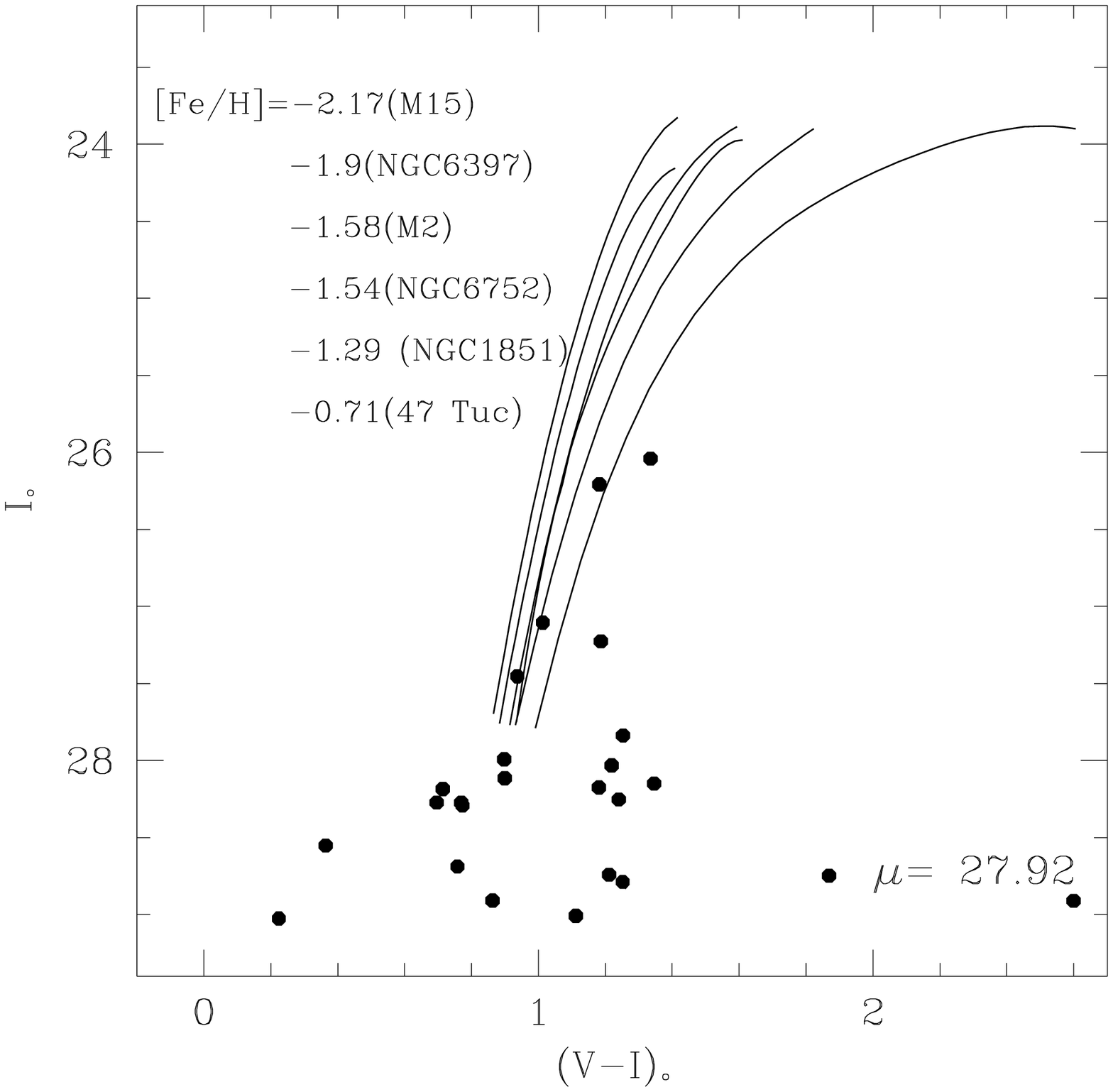}
\includegraphics[angle=0,width=0.45\textwidth]{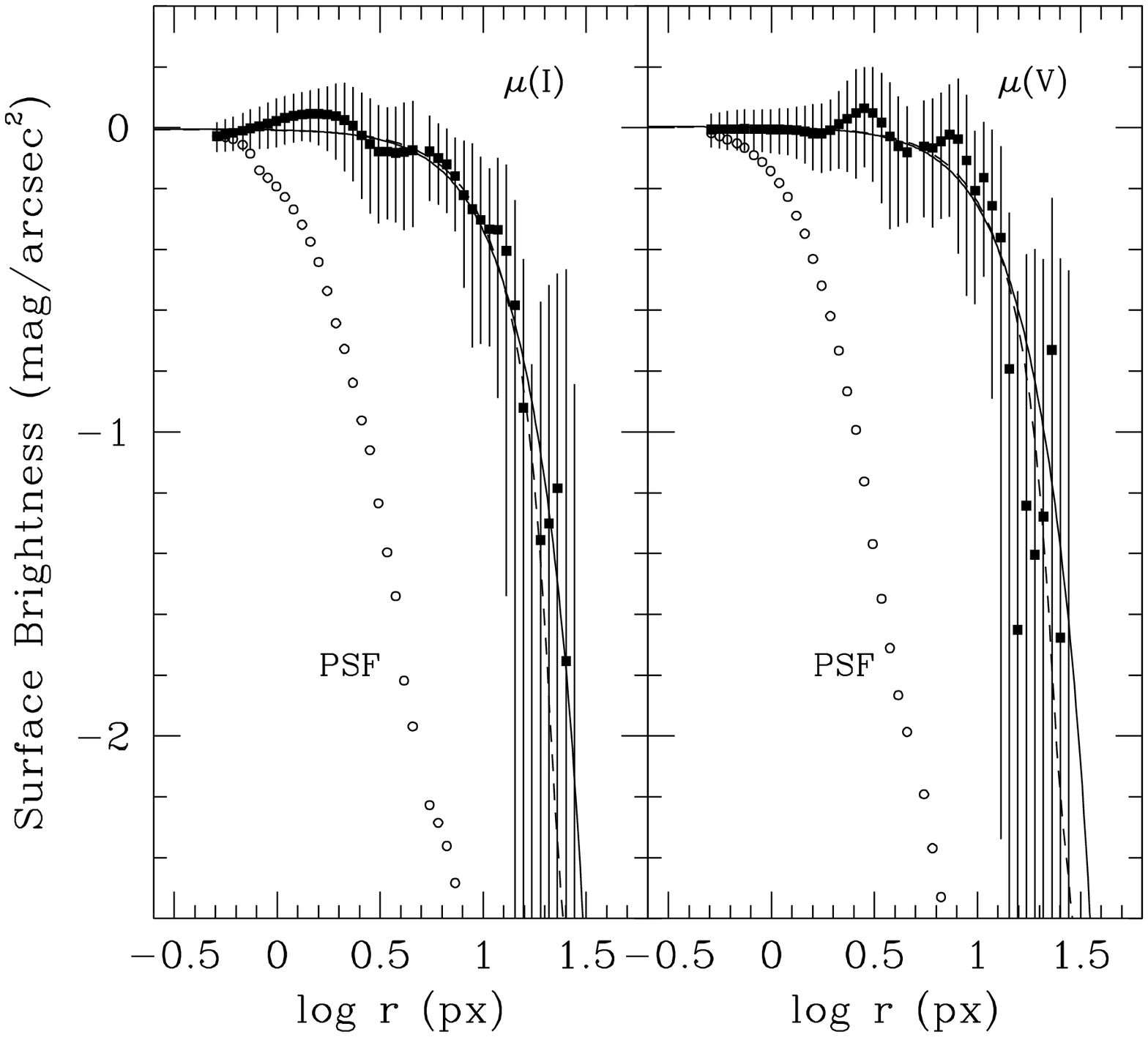}
\caption{Left: The colour-magnitude diagram of stars lying within a radius of 2 arcsec from 
the centre of the faint cluster GC0607. Shown also as solid lines are the red giant sequences 
of the indicated Galactic globular clusters from \citet{dacosta90}. Right: Surface brightness 
profile for the fainter of the two cluster candidates, GC0607, measured with ellipse-fitting as 
described in the text.  Symbols and lines are the same as in Fig.\,\ref{brightprofile}.}
\label{faintcluster}
\end{figure*}

\subsection{A second extended cluster candidate}

Motivated by the discovery of the candidate GC0606, we visually examined more closely  
the $V$ and $I$ master images looking for even fainter objects that could plausibly be 
Palomar-like GC candidates. A candidate located at $\alpha = 13^{\rm h}25^{\rm m}19.78^{\rm s}$, 
$\delta = -43\degr 34\arcmin 40.30\arcsec$ was found, which we label as GC0607. 
The middle panel of Figure\,\ref{images_gc} shows a zoom on the vicinity of this object.  
This second candidate draws attention to itself primarily as a symmetric, low-surface-brightness 
patch that is only marginally resolved into a few stars; it has few or no obvious bright RGB 
stars that would have allowed it to be picked up through our initial search technique 
(Fig.~\ref{xy_map}). The left panel of Fig.\,\ref{faintcluster} shows the CMD of all stellar 
objects detected within a circular region of radius of two arcminutes around the star cluster 
candidate. RGB fiducials are overplotted as before. No clear sequence of RGB stars is present, 
but the CMD contains a few stellar objects with luminosities and colours consistent with those 
expected for horizontal branch stars at the distance of NGC~5128. For comparison, the handful 
of faintest known GCs in the Milky Way such as Pal 13, Pal 1, or AM4 (all of which lie at 
$M_V < -4$ integrated luminosity) are almost entirely lacking in RGB stars brighter than the 
horizontal branch, to such an extent that the addition or subtraction of just one RGB star 
would change the total cluster luminosity very noticeably.

To determine the structural parameters of the cluster candidate, which is (mostly) unresolved,
we used \emph{stsdas/ellipse} to construct the empirical radial profile of light intensity out 
to a radius of 30 px ($1.5''$). Note that the circular aperture photometry leads to the same 
results to within the errors. As before, the comparison profile of the stellar PSF was measured 
from adjacent bright, unsaturated, uncrowded stars. The profile fitting code of \citet{mcl08} 
was then used to fit \citet{king62} and \citet{king66} profiles, numerically convolved with the 
PSF. The best-fit results are shown in the right panel of 
Figure \ref{faintcluster}, with the measured parameters listed in Table \ref{list_gc_candidates}. 
As for the first extended cluster, GC0607 is almost entirely dominated by a large, flat central 
core. Again, the central concentration and central dimensionless potential are similar to the 
lowest-concentration globular clusters known in the Milky Way. The half-light radius, the central 
surface brightness, and the total luminosity in each photometric band were determined in the 
same manner as for the first faint extended globular cluster. The difference in $(V-I)$ colour 
between GC0606 and 0607 is surprisingly large, but can be at least partly attributed to the 
relative dominance (or lack) of RGB stars brighter than the horizontal branch. GC0606 has 
numerous RGB stars, as seen from the colour-magnitude diagram, while GC0607 has virtually 
none and so its colour (if it is indeed a GC) would be dominated by the bluer subgiant and 
turnoff stars. 

The integrated luminosity, colour, and scale size of GC0607 are all consistent with it being 
an extremely faint, diffuse GC.  However, further and more definitive confirmation (such as 
by radial velocity measurement) will be exceptionally hard to obtain. Although we suggest 
that GC0607 should be kept as a plausible cluster candidate, the possibility cannot be ruled 
out that it is a faint background galaxy.


\begin{figure}
\includegraphics[clip=,width=0.45\textwidth]{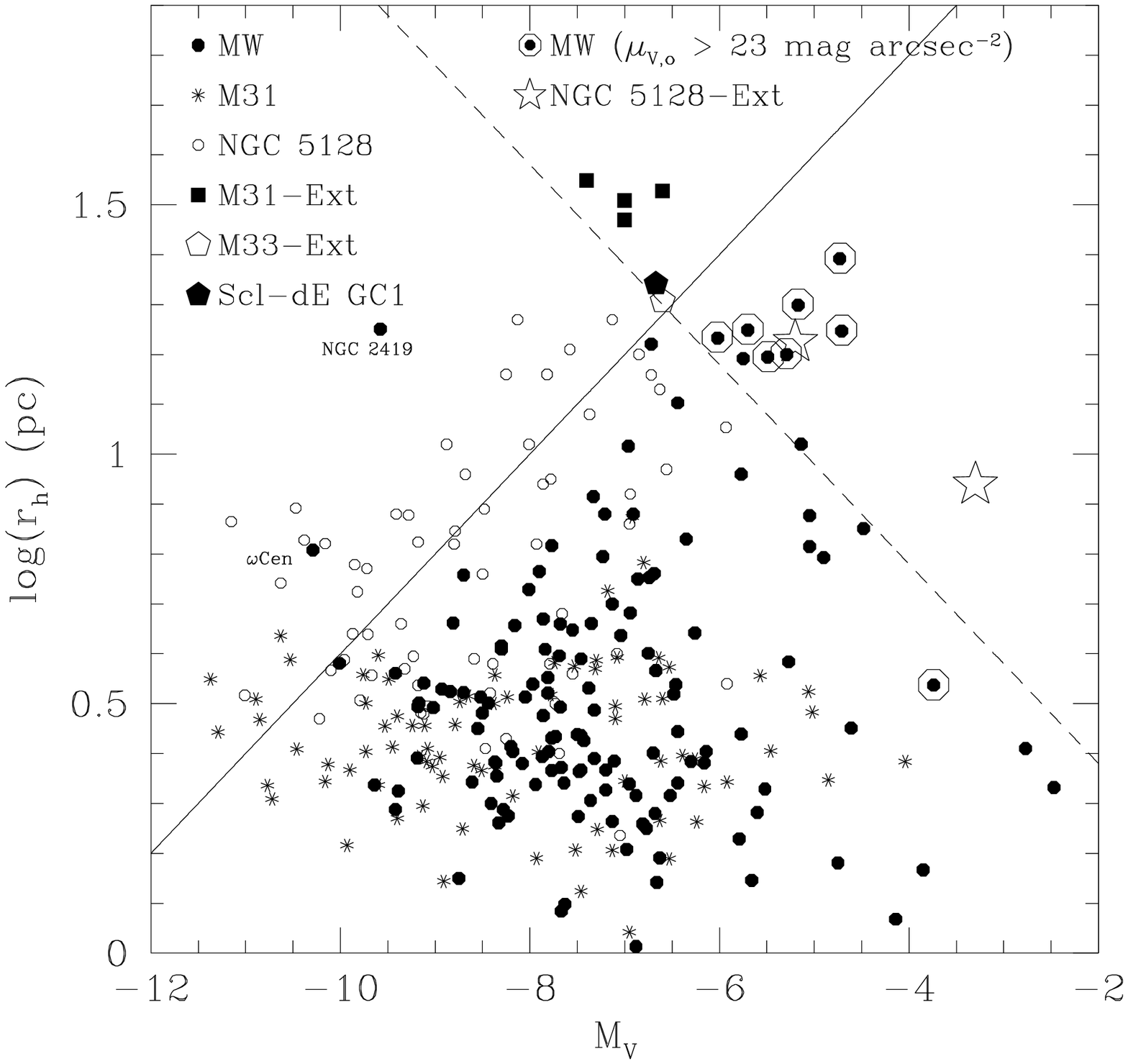}
\caption{Half-light radius (in pc) versus V-band luminosity for globular clusters in nearby 
galaxies. The Milky Way GCs from \citet{harris96} are shown as solid dots. Those with the 
central surface brightness values larger than ${\rm 23 mag\,arcsec^{-2}}$ are encircled. 
GCs of M31 \citep{barmby07} and NGC~5128 \citet{gomez06,mcl08} are shown as asterisks and 
open circles respectively. M31 extended clusters from \citep{mackey06} are shown as solid 
squares. Open and solid pentagons show respectively the extended clusters of 
M33 \citep{stonkute08}, and Scl-dE \citep{dacosta09}. The extended clusters discussed in the 
paper are shown as open stars. The solid line shows the relation $\log(r_h)=0.2\times M_V + 2.6$ 
from \citet{vandenbergh04}, and the dashed line represents a constant surface luminosity of
${\rm 15\,L_{V,\odot} \times pc^{-2}}$ within the half-light radius. The Galaxy GC $\omega$ Cen
and NGC 2419 are labeled.}
\label{rhmv}
\end{figure}


\begin{figure}
\includegraphics[clip=,width=0.45\textwidth]{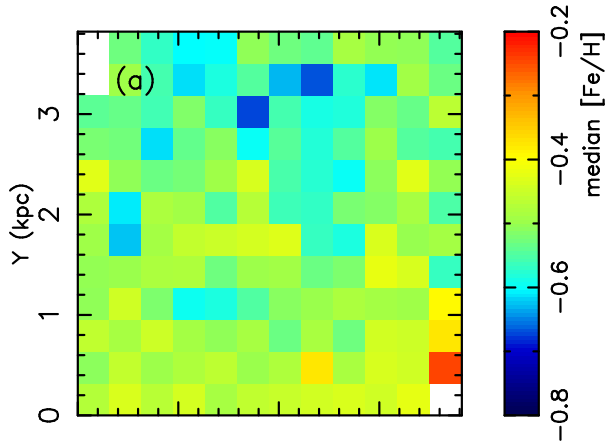}
\caption{Two-dimensional median metallicity map over the observed ACS field. 
Stars shown in Fig.~\ref{xy_map} binned into a $12\times12$ super-pixel array. }
\label{map_median_feh}
\end{figure}


\section{Discussion}
\label{disc}

Fig.\,\ref{rhmv} shows the location of the two new GC candidates we have discovered here 
on the structural parameter plane of half-light radius versus V-band luminosity, compared 
to those of the Milky Way \citep{harris96}, M31 \citep{barmby07} and NGC~5128 
\citep{gomez06,mcl08} shown as solid dots, asterisks, and open circles respectively. 
Galactic GCs with V-band extinction-corrected central surface brightness values fainter 
than ${\rm 23\,mag\,arcsec^{-2}}$ are encircled. The extended clusters reported around 
M31 \citep{mackey06}, M33 \citep{stonkute08}, and the Sculptor Group dwarf elliptical 
Scl-dE1 \citep{dacosta09} are also shown. The solid line shows the equation derived by 
\citet{vandenbergh04} who found that only two GCs lie above this line. A constant surface 
luminosity of ${\rm 15\,L_{V,\odot} \times pc^{-2}}$ within the half-light radius is also shown 
as the dashed line. This line seperates nicely the subpopulation of extended clusters from 
the rest of the GC population. 

The GC candidates we found in NGC~5128 appear different from the populations of extended 
clusters recently reported in the literature. Those objects are bright, i.e., $M_{V,\circ} \la -6.5$, 
while the extended clusters of NGC~5128 are much fainter matching up well, particularly the 
G0606 cluster, with the faint and extended cluster population in the Milky Way. 
Interestingly, the group of Galactic GCs with luminosities and half-light radii comparable to 
those of G0606 are situated all at galacto-centric distances $\ga 25\,$kpc. Only one Galactic 
GC beyond 25\,kpc, i.e., Palomar 13, is not in this group of clusters. It is worth to mention that 
the central surface brightness of the object of the group of Galactic GCs with luminosities and 
half-light radii similar to those of G0606 that is not encircled, i.e., Pyxis, is not available.

By examining the distribution of old globular clusters in a number of different galactic systems 
in the size vs. luminosity diagram, \citet{dacosta09} have argued that the distribution function 
of globular cluster sizes is most likely bimodal. A first dominant subpopulation has half-light 
radii peaking at $\sim3.0\,$pc, but there is a secondary subpopulation with half-light radii 
peaking at $\sim10\,$pc and a dearth of GCs with half-light radii around 5\,pc. 
They suggested that this indicates the presence of two distinct modes of star cluster formation, 
with the less common extended clusters being primarily formed in the gravitationally smoother 
environment of dwarf galaxies compared to larger galaxies. 

An intriguing implication of this correlation is that extended and diffuse GCs in large galaxies, 
which tend to be found predominantly in their outer regions \citet{vandenbergh04,huxor05,huxor09}, 
may be of accretion origin. By investigating the distributions of the Milky Way globular cluster 
properties, their spatial distribution, and by comparing with those of globular clusters in dwarf 
galaxies, \citet{vandenbergh04} have suggested indeed that the faint extended globular clusters 
in the outer Galactic halo originate most likely from the disruption of now defunct dwarf companions. 
Could the diffuse and under-luminous globular cluster we have discovered in the outskirts of 
NGC~5128 be the counterpart of the potentially accreted Galactic outer halo extended clusters?

If a star cluster is of accretion origin, one would expect, if the accretion event is not too old 
and/or the accreted stars are not fully mixed yet, that the diffuse stellar populations in its close 
vicinity would be different from the average halo stellar populations. The depth of our ACS 
imaging data resolving the stellar content of the halo down to the core helium burning stars 
gives us the opportunity to investigate the properties of the diffuse stellar populations in the 
immediate neighborhood of the extended globular cluster.

The ACS image contains many large background galaxies, as well as several bright stars (see 
Fig.\,1 of \citet{rejkuba05}) whose ``halos'' cause localised holes in the star-counts map. 
A mask was constructed by choosing suitably large elliptical areas around these problematic 
regions. We have binned the stars in a $12\times12$ grid in order for each super-pixel to 
contain $\sim 100$ sources (to have signal to noise ratio S/N$\sim 10$). Each superpixel is 
thus $17''$ wide or 314 parsecs projected width. Only stars outside of the masked regions 
were kept. A detailed and quantitative analysis of the spatial sub-structures over the ACS field 
is deferred to a subsequent contribution, but it is worth mentioning here that the stellar surface 
density of RGB stars over the field shows no obvious coherent stellar structure that might 
indicate the presence of streams or shells of stellar debris. 
Figure \ref{map_median_feh} shows the two-dimensional map of the median values of metallicity 
of RGB stars calculated in the super-pixel grid. For reference, the extended globular cluster 
discussed in the present paper is located within the most metal-poor super-pixel, the dark 
blue spot of Figure \ref{map_median_feh}. The random errors on the median metallicities in 
the super-pixels, due to both photometric uncertainties and population sampling effects, 
estimated as described in full details in \citet{ibata09}, are small, i.e., typically $\sim 0.02$~dex. 

A striking feature of the median stellar metallicity map is the large pixel-to-pixel variation, 
much larger than the typical random uncertainties. It is worth noting that those small-scale 
chemical variations have no obvious correspondence in the stellar density map. 
Due to the virtually negligible contamination from foreground or background sources, 
the present observations are extremely sensitive to the presence of sub-structures. 
It is in principle possible with the ACS to detect a population of $\sim 15$ RGB stars scattered 
over a volume of many hundred cubic kpc, making it much easier to detect sub-structures from 
spatial metallicity variations rather than from the enhancement they cause in the stellar density 
map \citep[see][for more details]{ibata09}.

The outer regions of NGC~5128 have been found to be predominantly populated by 
moderately metal-rich stars, with an average metallicity of ${\rm [m/H]\sim -0.5}$ and 
only $\sim 10\%$ of the stellar populations more metal-poor than ${\rm [m/H]\approx -1}$ 
\citep{harris00,harris02,rejkuba05}. A gradient of the median metallicity is apparent in the 
median metallicity map, with the top half of the ACS field being more metal-poor than the 
lower half. As discussed at length by \citet{rejkuba05}, the metallicity distribution function 
of the overall stellar populations at $\sim 40$\,kpc is very similar in term of its shape, average 
metallicity, and metallicity dispersion, to those determined at $\sim 20$ and $\sim 30$ kpc 
respectively from the centre of the galaxy \citep{harris99,harris00}. No detectable radial 
metallicity gradient is present in the outskirts of NGC~5128, at least in the range of radial 
distances probed. The detected median metallicity gradient within the ACS field at 
$\sim 40$\,kpc from the centre of the galaxy should thus not reflect a global metallicity 
gradient, which would give unreasonably high metallicities if extrapolated inward. 
The median stellar metallicity gradient is most likely related to localised chemical variations 
that have not yet been fully blended into the diffuse smooth halo. The spatial correspondence 
and the similar metallicities between the small-scale chemical variations and the extended 
globular cluster suggest that this object is likely to be related to the same event that has led 
to the formation of the localised chemical variations.

\section{Summary}
\label{summary}

In this paper, we have presented a discovery and discussion of two new globular cluster candidates 
in the outer region  ($d \sim 40\,$kpc from the centre) of the nearest giant elliptical NGC~5128 
using very deep HST/ACS imagery. The first and brightest of these two objects is selected based 
on both the presence of a clear RGB sequence and its structural parameters. The second one was 
selected solely on basis of its morphological appearance, colour and luminosity, and its structural 
parameters. The properties of the bright cluster are consistent with its identification as an old, 
intermediate-metallicity globular clusters resembling in every way we can verify the faint and 
extended clusters populating the Milky Way outer halo. This is the first time a counterpart of 
this populations of Galactic globular clusters has been reported for any other galaxy. 
The combined properties of the extended clusters and the diffuse stellar halo are consistent 
with the view that the reported clusters were once associated with dwarf galaxies that have 
disrupted.

Finally, our present work indicates that a search for faint, extended globular clusters in 
NGC 5128 is entirely feasible. By carefully searching just one deep ACS/WFC imaging 
field -- which in this context is essentially a random pointing in the halo -- we have 
successfully identified two very probable GC candidates and characterized their structural 
parameters. Extrapolating from just this one field of area 11.5 arcmin$^2$ to the entire 
NGC 5128 halo area would be extremely risky, but it is worth noting that the Milky Way has more 
than a dozen known GCs fainter than $M_V \simeq -5$ \citep{harris96}. NGC 5128 is a more luminous 
galaxy by an order of magnitude, so it seems quite likely that its vast outer halo should hold 
some hundreds of faint GCs awaiting discovery, along with their attendant substructures.

\section*{Acknowledgments} 

MM thanks Gary Da Costa for providing some of the data shown in Fig. 8 in an electronic
form. This research has made use of the SIMBAD database, operated at CDS, Strasbourg, France.

\bsp

\label{lastpage}

\end{document}